# Efficacy of EPSS in High Severity CVEs found in CISA KEV

Rianna Parla
Independent Research
rparla@alumni.purdue.edu

***Abstract*— The Exploit Prediction Scoring System (EPSS) is designed to assess the probability of a vulnerability being exploited in the next 30 days relative to other vulnerabilities. The latest version, based on a research paper published in arXiv [1], assists defenders in deciding which vulnerabilities to prioritize for remediation. This study evaluates EPSS's ability to predict exploitation before vulnerabilities are actively compromised, focusing on high severity CVEs that are known to have been exploited and included in the CISA KEV catalog. By analyzing EPSS score history, the availability and simplicity of exploits, the system's purpose, its value as a target for Threat Actors (TAs), this paper examines EPSS's potential and identifies areas for improvement.**

## Introduction

### A. What is EPSS, CVSS, and CVE?

The Exploit Prediction Scoring System (EPSS) is a well-established machine-learning based scoring system used to estimate the likelihood of a security vulnerability being exploited within the next 30 days. EPSS is used alongside the Common Vulnerability Scoring System (CVSS), which rates the severity of vulnerabilities from 0 to 10 based on potential damage. Together, these systems are vital for cyber-security experts, as it enables them to prioritize vulnerability remediation effectively [2]. Common Vulnerabilities and Exposures (CVE) [3] is a system that categorizes and classifies security vulnerabilities. It uses the Common Vulnerability Scoring System (CVSS) [4] to evaluate the threat level of each identified vulnerability. While EPSS can help detect vulnerabilities likely to be exploited, further research and discussion is needed to see if it is truly effective in predicting high-severity vulnerabilities before they are widely exploited.

### B. What is CISA KEV?

The Known Exploited Vulnerabilities (KEV) catalog, maintained by the Cybersecurity and Infrastructure Security Agency (CISA), lists vulnerabilities that have been confirmed as actively exploited in the wild. These vulnerabilities represent a subset of CVEs (Common Vulnerabilities and Exposures), which are represented by a CVE ID. The US Government issued a binding directive to ensure that vulnerabilities in the CISA KEV catalog were addressed by organizations doing business with the Federal Government. To be included in the CISA KEV catalog, the following criteria must be met [3]:

1) Active exploitation of the vulnerability has been observed, with either successful attacks or attempts that specifically target the vulnerability
2) Clear remediation guidance must be available.

### C. Why is prioritizing vulnerabilities important?

At the time a vulnerability has been included in the CISA KEV catalog, it is already widely known to have been exploited. Due to the sheer volume of vulnerabilities, fixing every single one in a timely manner is simply not possible. On average, companies worldwide take between 88 and 208 days to patch vulnerabilities [5], so it is important to prioritize the solutions for the most dangerous (i.e., most likely to be exploited) ones first.

### D. Why study this topic?

The purpose of this study is to assess if EPSS scores are truly predictive rather than trailing indicators, as well as how useful they are for cybersecurity defenders. Although EPSS scores can detect vulnerabilities that lead to real exploits, there is a lack of research on whether or not the system is predictive. By assessing data from high severity CVEs, this study aims to validate the effectiveness of EPSS as a tool for vulnerability management, and to identify and discuss potential areas of improvement.

## Background

### A. EPSS system overview

The Exploit Prediction Scoring System (EPSS) is a data-driven solution that uses machine learning to estimate the probability between 0 and 1 (0 to 100%) that a software vulnerability will be exploited in the wild relative to other vulnerabilities. This is used alongside other systems, such as CVSS, which provides a score from 0 to 10 based on the severity of danger that the vulnerability poses if it were to be exploited [2]. All of this data is publicly available and essentially, CVSS measures how much damage a vulnerability could cause if exploited, and EPSS measures the likelihood that a vulnerability will be exploited. The goal of these systems is to assist network and other threat defenders to better prioritize vulnerability remediation efforts, and when used together, EPSS and CVSS provide a more comprehensible approach to these efforts.

### B. How are EPSS scores calculated?

EPSS focuses on vulnerability prediction and uses current threat information from CVEs and real-world exploit data [6]. The authors of the EPSS system indicate that they are using the gradient-boosted trees (XGBoost) machine learning model [7]. A third iteration of their system boasts improved training of the XGBoost model.

Note: EPSS v3 remediation strategy would prioritize vulnerabilities with EPSS probabilities of 0.36 (36%) and above. [1]

TABLE I: Description of data sources used in EPSS [1]

| Description | Variables | Type | Sources |
|---|---|---|---|
| Exploitation activity in the wild (labels) | 1 (with dates) | Binary | Fortinet, AlienVault, Shadowserver, GreyNoise |
| Publicly available exploit code | 3 | Binary | Exploit-DB, GitHub, MetaSploit |
| CVE mentioned on list or website | 3 | Binary | CISA KEV, Google Project Zero, Trend Micro ZDI |
| Social media | 3 | Numeric | Mentions/discussion on Twitter |
| Offensive security tools and scanners | 4 | Binary | Intrigue, sn1per, jaeles, nuclei |
| References with labels | 17 | Numeric | MITRE CVE List, NVD |
| Keyword description of vulnerability | 147 | Binary | Text description in MITRE CVE List |
| CVSS metrics | 15 | One-Hot | National Vulnerability Database (NVD) |
| CWE | 188 | Binary | National Vulnerability Database (NVD) |
| Vendor labels | 1,096 | Binary | National Vulnerability Database (NVD) |
| Age of the vulnerability | 1 | Numeric | Days since CVE published in MITRE CVE list |

In Table I, there are four columns representing description, variables, type, and sources. The variables column represents the number of variables for that description type. For example, social media has 3 variables associated with that data type. The type column indicates the data representation format. Each row describes the individual elements used as label data in the machine learning algorithm. While Table I outlines certain variables used as data features, they do not detail the exact data features used within the model. For example, GreyNoise serves as a threat feed, but the specific data features extracted from that threat feed are not specified. The EPSS system leverages a total of 1477 features to predict exploitation activity. The precise features being used are not documented in their paper.

## Methodology And Assumptions

This research methodology relies exclusively on publicly available data to evaluate the efficacy of the Exploit Prediction Scoring System (EPSS). The dataset is primarily composed of Common Vulnerabilities and Exposures (CVE) data sources, EPSS scores of these CVEs, and the Cybersecurity and Infrastructure Security Agency (CISA) Known Exploited Vulnerabilities (KEV) catalog. Other data sources include relevant news articles, available exploits, proof-of-concept samples, and and any other public data sources that may be useful for vulnerability analysis.

The dataset will be limited to the introduction of the latest EPSS scoring system (EPSSv3) that was deployed in March 2023 [6].

While a CVE being evaluated may predate the latest EPSS ecosystem, it will only be considered if it was added to the CISA KEV catalog after 03/15/2023. This ensures that the EPSSv3 prediction system would have been active prior to addition of that CVE into the CISA KEV catalog.

As part of the analysis, particular attention will be given to CVEs with a score of 9.0 or higher, as these represent the most serious vulnerabilities available. As of the time of this research, there were 90 CVEs that made the CISA KEV catalog with a score of 9.0 or higher between 03/15/2023 through 07/01/2024. The research will focus on vulnerabilities that do not require authentication in systems that present the most potential value to a threat actor (TA), such as a remote access system, where a compromise would lead to a large set of internal resources that might also be targeted. Similarly, popular software used in internet facing systems will be closely analyzed given the likelihood of exploitation, as will vulnerabilities known to have been used in ransomware campaigns [8].

Several assumptions underlie this research. They will be outlined here and further explored in the sections directly below.

1) The EPSS system should be able to reasonably predict the likelihood of a CVE being exploited before it is listed in the CISA KEV catalog. It is assumed that EPSS scores correlate with the presence of a CVE in the CISA KEV catalog, except in cases where a vulnerability is disclosed and immediately added to the catalog.
2) A CVE's presence in the CISA KEV catalog implies it has been successfully exploited, or was known to be specifically targeted to be exploited as reported by CISA. The research assumes that vulnerabilities in the catalog represent those that should be prioritized for mitigation by organizations.
3) Remote access technologies, VPNs, and perimeter security systems are critical areas of vulnerability, and are considered high-value targets for TAs. This assumption is based on their frequent presence in the CISA KEV catalog.

### A. Presence in CISA KEV

As mentioned above, a vulnerability that is listed in CISA KEV has either been successfully exploited, or was known to be targeted for exploitation. Research has shown that on average, an exploit is published about 3 weeks before a CVE is disclosed to the public [9]. Furthermore, a vulnerability may exist long before that time due to responsible disclosure practices. A patch or other mitigation must be present before a CVE is added to the CISA KEV catalog, which suggests that a vulnerability can exist for a significant period of time before the CISA KEV catalog contains the CVE associated with a vulnerability. Recent studies have shown that the average time to fix a vulnerability can be 60 days or more [10].

### B. Expectations of EPSS prediction

The EPSS system is expected to gauge the likelihood of exploitation of a vulnerability before it is added to the CISA KEV catalog. Many vulnerabilities are well-known to the vendor and a small number of other entities for some period of time before the vulnerability is disclosed in a CVE. Oftentimes these vulnerabilities have been exploited on a "Patient Zero" long before a patch is available. Once a patch is available, a public disclosure in the form of a CVE is crafted and made generally available. EPSS should reflect this risk effectively, with scores correlating to the probability of exploitation. Exceptions to this assumption include instances where a CVE is immediately added to the CISA KEV catalog upon disclosure.

### C. *Remote Access Technologies as High Value Targets*

In January 2022, the White House released a directive (M-22-09) to move away from VPN to a Zero Trust (ZT) architecture, due to the ongoing exploitation of remote-access VPNs. This directive supports Executive Order 14028, which was made to improve the United State's cybersecurity [11] [12].

In early 2024, CISA published a joint recommendation to move to a more modern security implementation for access that was not VPN-centric [13]. In fact, CISA themselves were compromised by a VPN vulnerability, leading to a breach. That same VPN product had a series of high severity CVEs, many of which made the CISA KEV catalog [14]. Several advisories have been issued in 2024 related to VPNs by the US Government including a multi-agency joint cybersecurity advisory highlighting the serious nature of the threat [15].

Every major VPN, firewall, or perimeter security vendor has had at least one CISA KEV cataloged CVE in either 2023 or 2024, highlighting the critical importance of an effective exploit prediction system for this area of technology. Networking and perimeter systems, in particular, have had significant presence in the CISA KEV catalog over recent years compared to other technologies, warranting additional research and analysis. Among the 90+ CVEs with a CVSS score of 9.0 or higher listed in the CISA KEV catalog since March 2023, a notable 22 were related to a networking or perimeter security product.

## CVE EPSS Example Data

In this section, we will examine a few representative examples of high-risk CVEs, particularly those impacting remote access technologies and productivity software. These CVEs illustrate the vulnerabilities commonly targeted by threat actors and their respective exploit timelines. In the following sections, CVE EPSS Score details will be shown by graphs, where the X-axis consists of dates, and the Y-axis is a score scale from 0 to 1. The blue line indicates the EPSS score, and the yellow line indicates the percentile. Three arrows will also be present on the graphs: the cyan arrow indicates the date where a CVE identifier was created for the vulnerability, the red arrow indicates when the vulnerability was added to the CISA KEV catalog, and the purple arrow indicates the first known in-the-wild exploit for the vulnerability.

### A. *CVE-2023-3519 (remote access technology)*

Citrix NetScaler ADC and NetScaler Gateway Code Injection Vulnerability. NetScaler Gateway is a popular security appliance widely deployed in enterprises, typically with direct exposure to the internet. A breach of one of these devices could lead to significant exposure of private resources within the enterprise network. A detailed write up about this vulnerability in a two-part blog, titled *"Analysis of CVE-2023-3519 in Citrix ADC and NetScaler Gateway"* was available on 07/04/2023 and 07/24/2023 [16]. The first blog was prior to when the CVE was originally published and added to CISA KEV (07/19/2023). The second blog entry noted that another vulnerability associated with CVE-2023-3519 was the one being exploited in the wild. This was known to have been used in a ransomware attack [17]. A Metasploit module was made available on 08/03/2023.

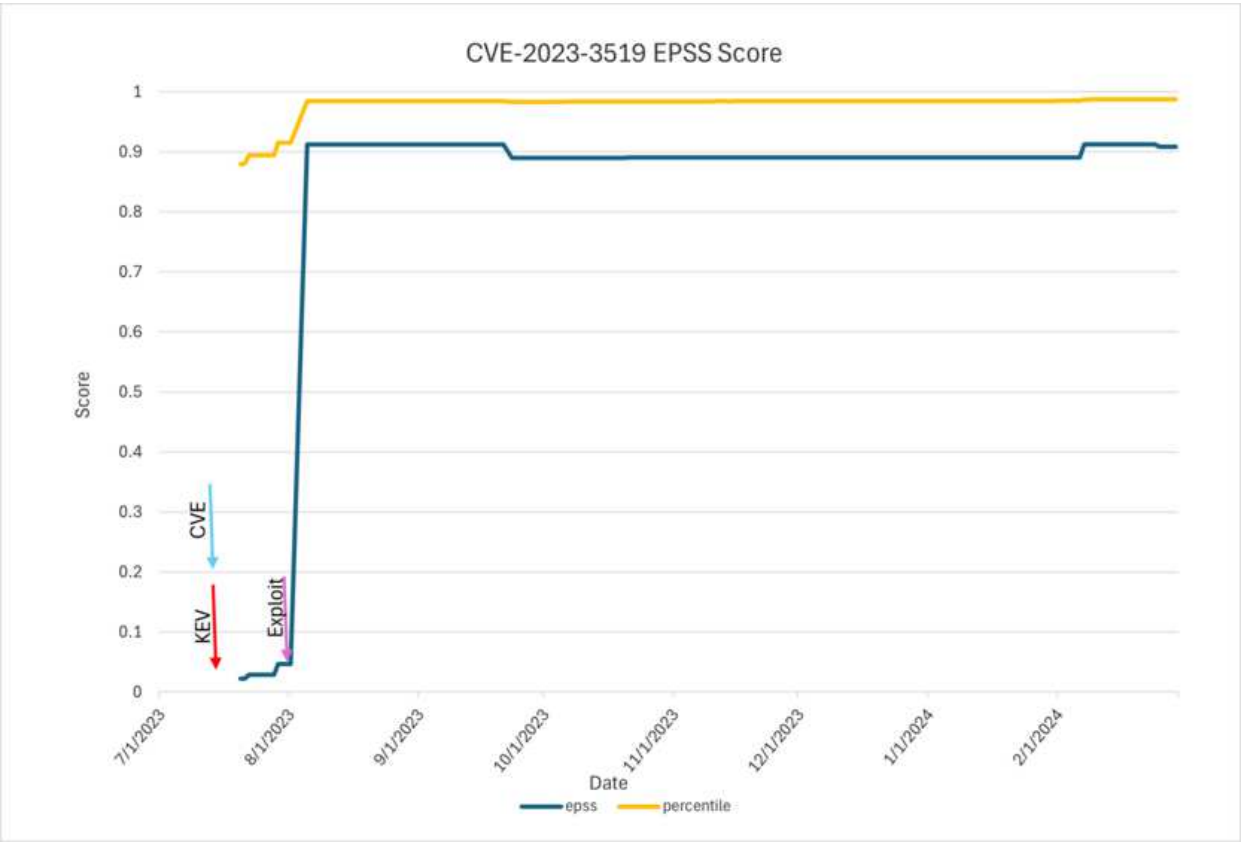

Fig. 1: CVE-2023-3519 EPSS Details

### B. *CVE-2023-4966 (remote access technology)*

Citrix NetScaler ADC and NetScaler Gateway Buffer Overflow Vulnerability. NetScaler Gateway is a popular security appliance widely deployed in enterprises, typically with direct exposure to the internet. A breach of one of these devices could lead to significant exposure of private resources within the enterprise network. A detailed write up about this vulnerability titled *"Citrix Bleed: Leaking Session Tokens with CVE-2023-4966"* was available on 10/24/2023 [18], shortly after the CVE was originally published (10/10/2023) and added to CISA KEV (10/18/2023). This was known to have been used in a ransomware attack [19]. A Metasploit module was made available on 10/31/2023.

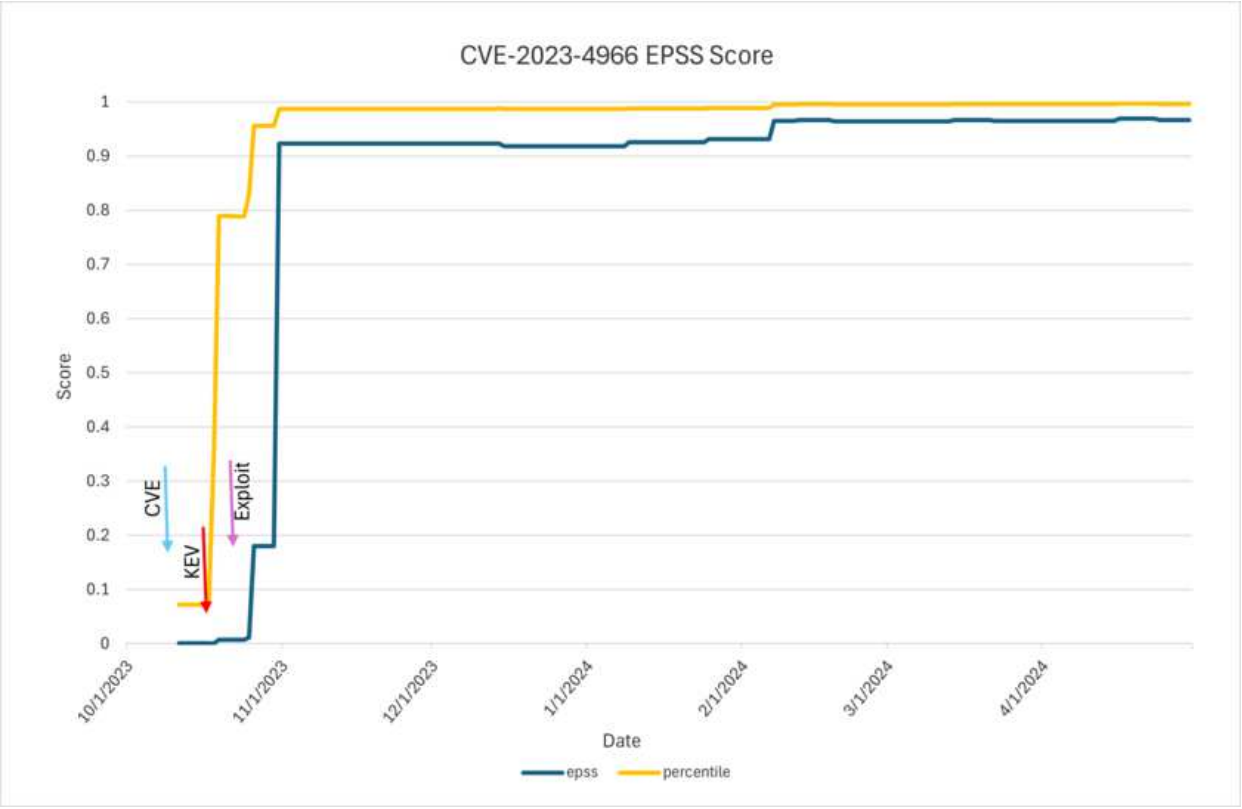

Fig. 2: CVE-2023-4966 EPSS Details

### C. *CVE-2023-7028 (supply chain software)*

GitLab password reset account takeover vulnerability. GitLab is a widely used open DevOps platform for source code management and builds. Taking over an account could result in a Supply Chain attack by a Threat Actor. A detailed write up of the exploit existed in December of 2023 [20]. Another detailed write up about this vulnerability titled *"Account Takeover via Password Reset without user interactions"* was published on 01/11/2024 [21], the same day the CVE was published. A public exploit was available the next day [22]. The CVE was not added to the CISA KEV catalog until 05/01/2024. A Metasploit module was made available on 03/07/2023.

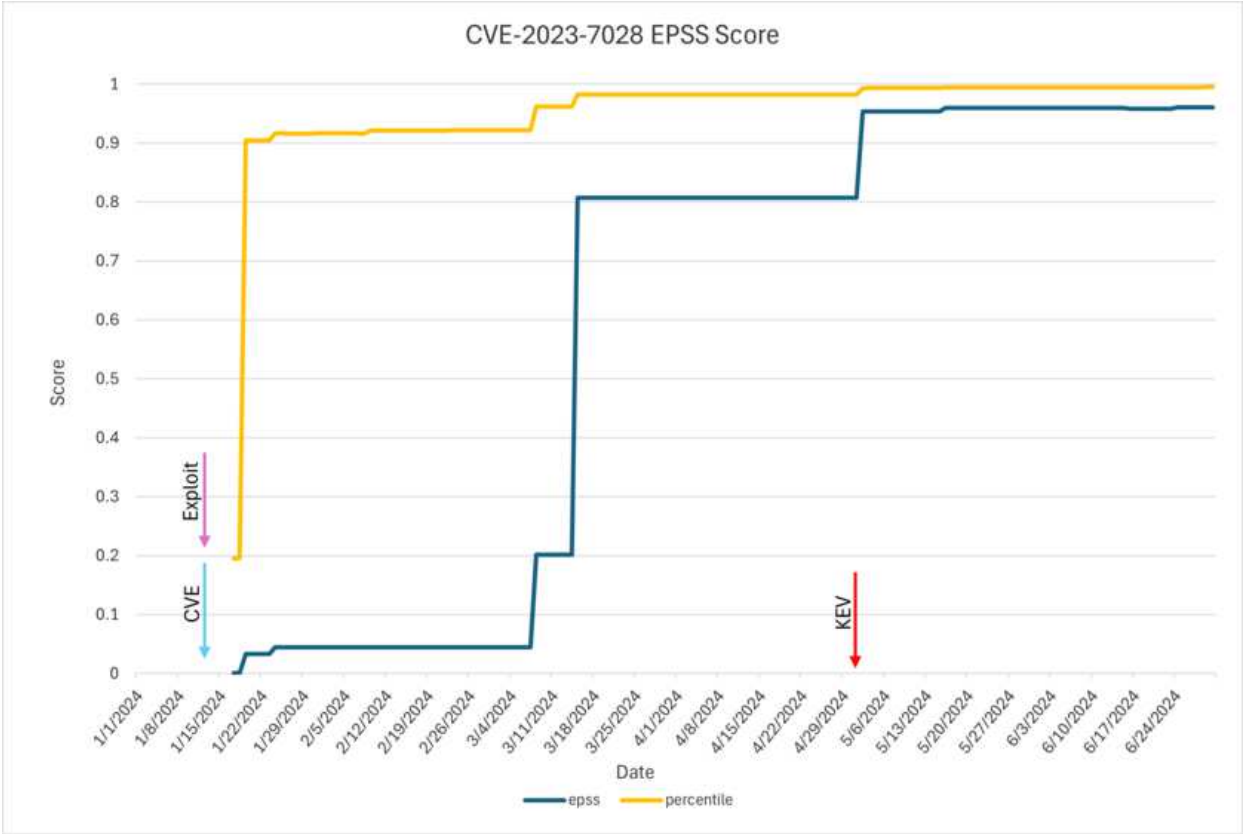

Fig. 3: CVE-2023-7028 EPSS Details

### D. *CVE-2023-22515 & CVE-2023-22518 (productivity software)*

Atlassian Confluence Data Center and Server Access Control Vulnerabilities allow threat actors to create Administrative accounts without authorization. Atlassian Confluence is a popular enterprise workspace for creating content. It is typically deployed internally, however, some smaller deployments are exposed to the internet, and this vulnerability was used in a ransomware attack [23]. Public exploits were available prior to when EPSS score exceeded the 36% threshold [24] [25]. Metasploit modules were made available on 10/19/2023, 10/20/2023 and 12/19/2023.

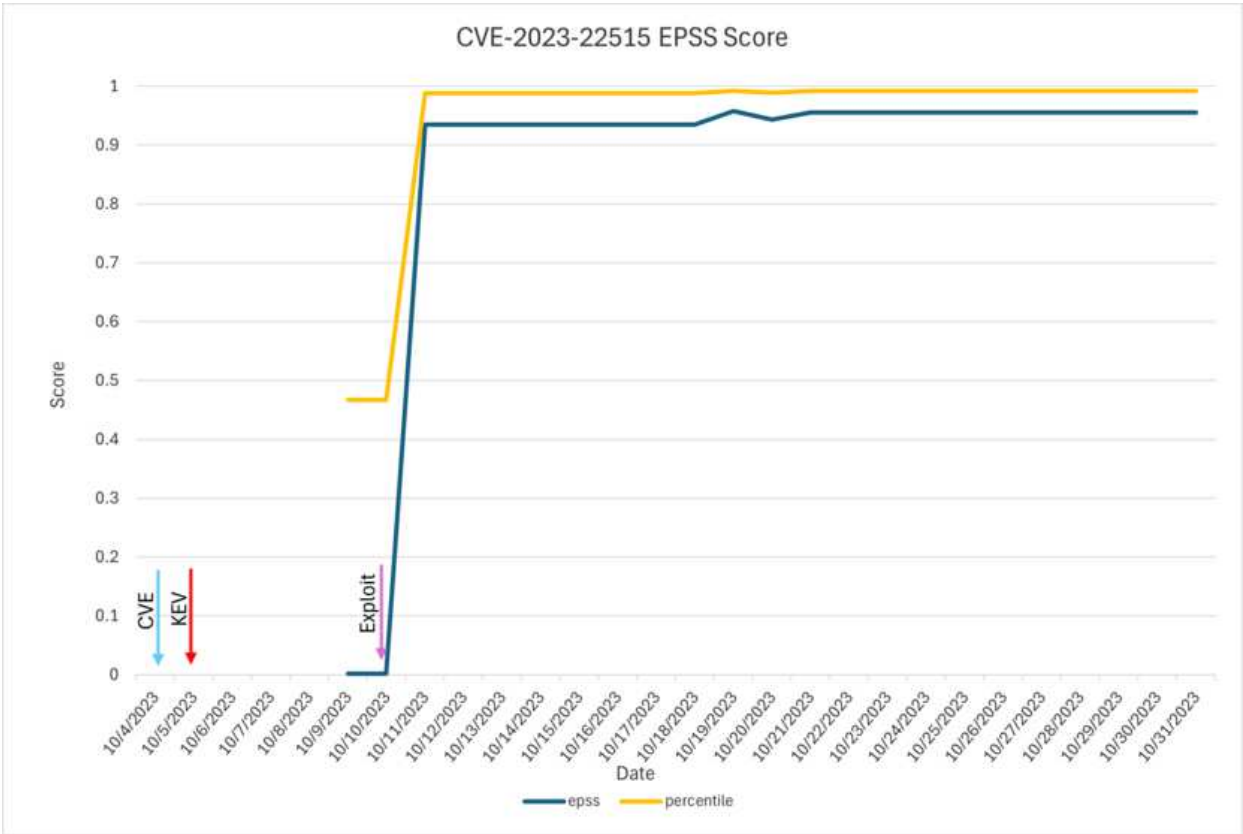

Fig. 4: CVE-2023-22515 EPSS Details

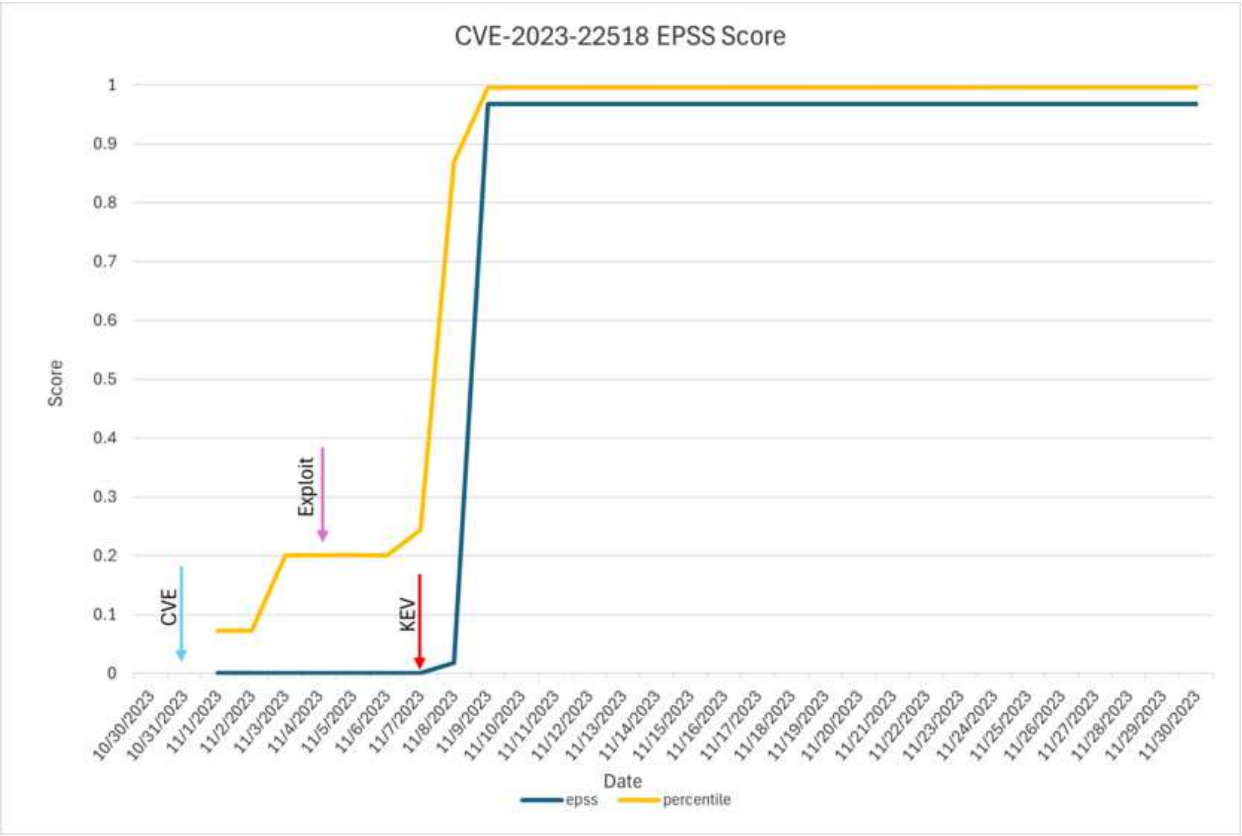

Fig. 5: CVE-2023-22518 EPSS Details

### E. *CVE-2023-22527 (productivity software)*

Atlassian Confluence Data Center and Server allows an unauthenticated attacker to achieve RCE. Atlassian Confluence is a popular enterprise workspace for creating content. It is typically deployed internally, however some smaller deployments are exposed to the internet, and this vulnerability was used in a ransomware attack [26]. A Metasploit module was made available on 01/26/2024.

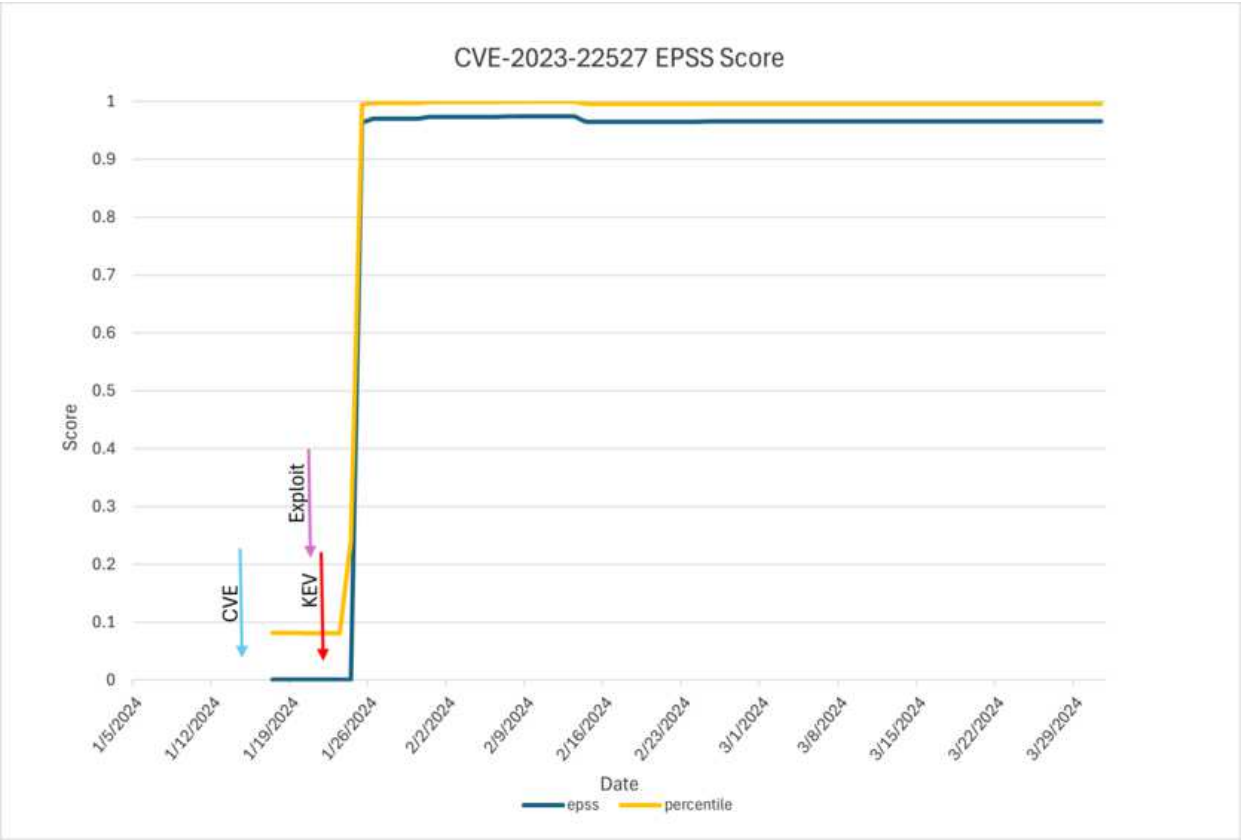

Fig. 6: CVE-2023-22527 EPSS Details

### F. *CVE-2023-27997 (remote access, perimeter security)*

Fortinet FortiOS and FortiProxy SSL-VPN Heap-Based Buffer Overflow Vulnerability. Fortinet is a popular network security vendor offering remote access and perimeter security appliances, typically with direct exposure to the internet. A breach of one of these devices could lead to significant exposure of private resources within the enterprise network. Fortinet is also a participant in the EPSS ecosystem providing telemetry data used by the EPSSv3 model.

bgImg

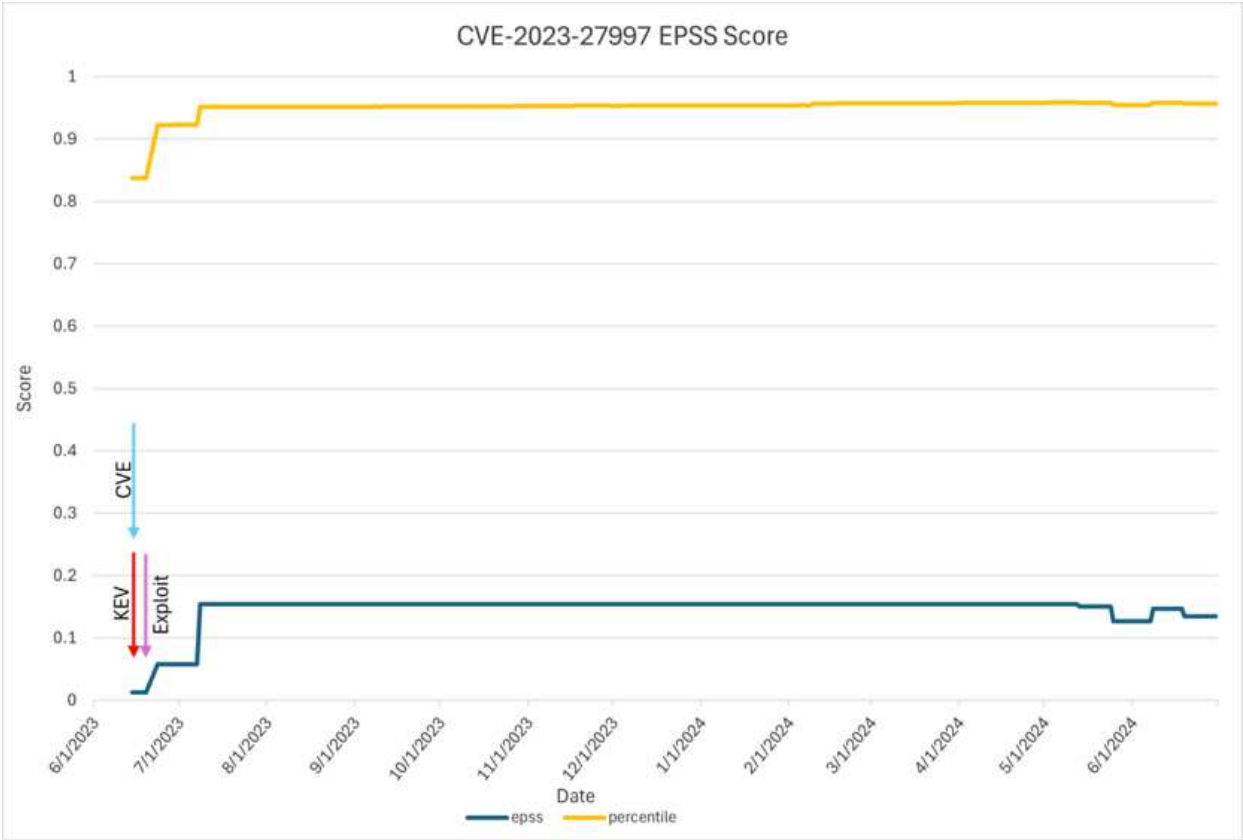

Fig. 7: CVE-2023-27997 EPSS Details

For additional examples and data on CVEs that follow similar patterns, refer to **Appendix A**.

## Discussion

Among the 90 CVEs meeting the research criteria, older CVEs tended to fair better in terms of having a high probability in the EPSS scoring system prior to CISA KEV inclusion. In fact, all but two CVEs that were from 2022 or earlier had high probability scores prior to inclusion in CISA KEV. For example, CVE-2020-17519 (Apache Flink) had a 90% probability score from the launch of the new EPSS Model in March 2023 up to the present and was added to the CISA KEV catalog in May 2024. However, the effectiveness of EPSS as a predictor remains unclear, as Palo Alto Networks Unit 42 reported extensive in-the-wild exploitation between November 2020 and January 2021 [27].

Conversely, newer CVEs examined tended to have mixed-to-poor results by comparison. Meaning if a CVE was created at or around the time of the introduction of the new EPSS model, they did not fair as well as a prediction system relative to CISA KEV. As noted, when a CVE is added to the CISA KEV catalog, it has been determined to have been exploited or attempted to be exploited already. CISA KEV is also a subset of all exploited CVEs and focuses on the most critical ones at the time of inclusion. Of the 57 CVEs from 2023 or later on the CISA KEV catalog, 42 had consistently low (mostly single digit) EPSS scores prior to inclusion in CISA KEV. 6 of the 57 CVEs had a score around 36% prior to inclusion in CISA KEV, and immediately jumped up on the day after inclusion in the CISA KEV catalog. 4 of the 57 were added to the CISA KEV catalog on the same day as the vulnerability was disclosed. All four had high EPSS scores on the initial release date, however were already in CISA KEV.

If one considers all 250 CVEs in CISA KEV during the research period, not just the ones with a CVSS score of 9.0 or greater, efficacy scores drop further for EPSS. Of the 250 CVEs, more than two thirds had a score below the 36% prioritization threshold just prior to being added to the CISA KEV catalog. Stated differently, only a third of the CVEs had an EPSS score reflective of it having been previously exploited and included in the CISA KEV catalog. This suggests that a prioritization strategy that focused on EPSS would leave a significant number of systems vulnerable for a period of time.

### A. *Model analysis*

There is limited public information on the efficacy of the models used by EPSS. One self-evaluation was done by Forum of Incident Response and Security Teams (First) on the Log4J vulnerability (CVE-2021-44228). The analysis looks at changes in the EPSS score over time and what event triggered that change. A detailed graph can be found on their website [28]. As the author notes in the blog, Log4shell was widely known about and did not need a prediction system to guide defenders into remediation. The EPSSv2 system was used for this evaluation and not the newer EPSSv3 system. As such, a newer CVE will need to be studied in a similar manner so that a comparison can be made as to the efficacy of the v3 model against a notable vulnerability. One item that stands out in the self-evaluation is how a single label feature (*Remote* tag removed) had resulted in a 10 point drop in the score. Similarly, the removal of a Metasploit module (penetration testing framework) from the public domain resulted in a score drop of more than 30 points. When the Metasploit module was republished, approximately 25 days later, the score returned to its previous value. During that window of time, a substantial number of new versions of public exploit permutations, including obfuscations, were being published in GitHub [29]. During this time, nothing about the vulnerability itself had changed.

A similar evaluation was done that looked at the volume of observed exploitation versus the EPSS score. It is clear in this evaluation that the EPSS score was primarily a trailing indicator of the large initial exploitation volume observed. In fact, according to the First evaluation of this data, described earlier, the initial jump was the result of the Metasploit module addition, not actual exploit observations. Even while the exploitation volume remained high, the EPSS score dropped sharply. Towards the latter portion of the graph, the EPSS score jumps back to its peak value while observed exploitation volume had not changed meaningfully. This suggests that other factors had a more meaningful impact on the model than the observed exploits [30]:

bgImg

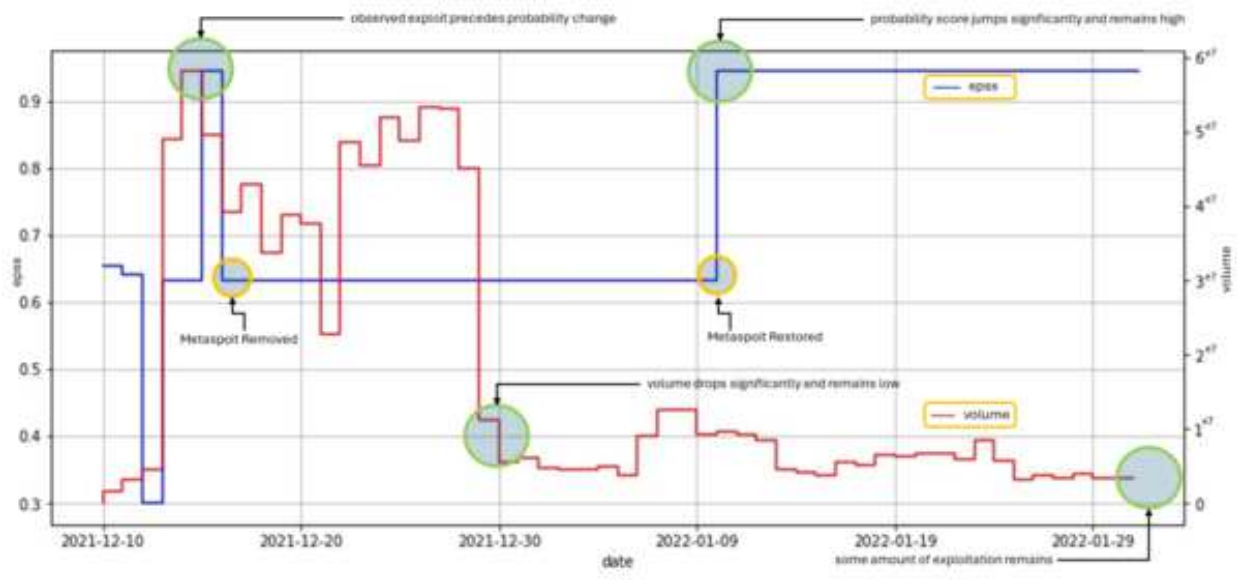

Fig. 8: Log4j EPSSv2 versus Exploit Volume (Fortinet)

### B. Deployment and Functional Considerations

How a system is deployed will affect the probably of exploitation. For example, a system directly exposed to the internet is significantly more at risk of exploitation than a system deployed in a private closed network. While the EPSS system cannot know what the deployment model is, it could hypothesize a score based on the the deployment model. For example, 3 scores could be provided showing the change in probability if exposed directly to the internet, on a private open network or on a private closed network.

Additionally, the role or purpose of the system does not seem to be a consideration in the modeling. For example, a remote access or perimeter security system would have a high probability of exploitation given the nature of that system. Similarly, a supply chain software ecosystem vulnerabilities would be more likely to be exploited because it would allow a Threat Actor to potentially inject *backdoors* or other vulnerabilities into other software systems. One improvement that should be considered in the modeling is to add this attribution as a data feature for consideration.

### C. CVE-2024-4040 (CrushFTP) analysis

This CVE is representative of the majority of CVEs evaluated during this research. It shows a significant lag between the time the CVE is added to the CISA KEV catalog and the EPSS score reflects a probability of exploitation:

bgImg

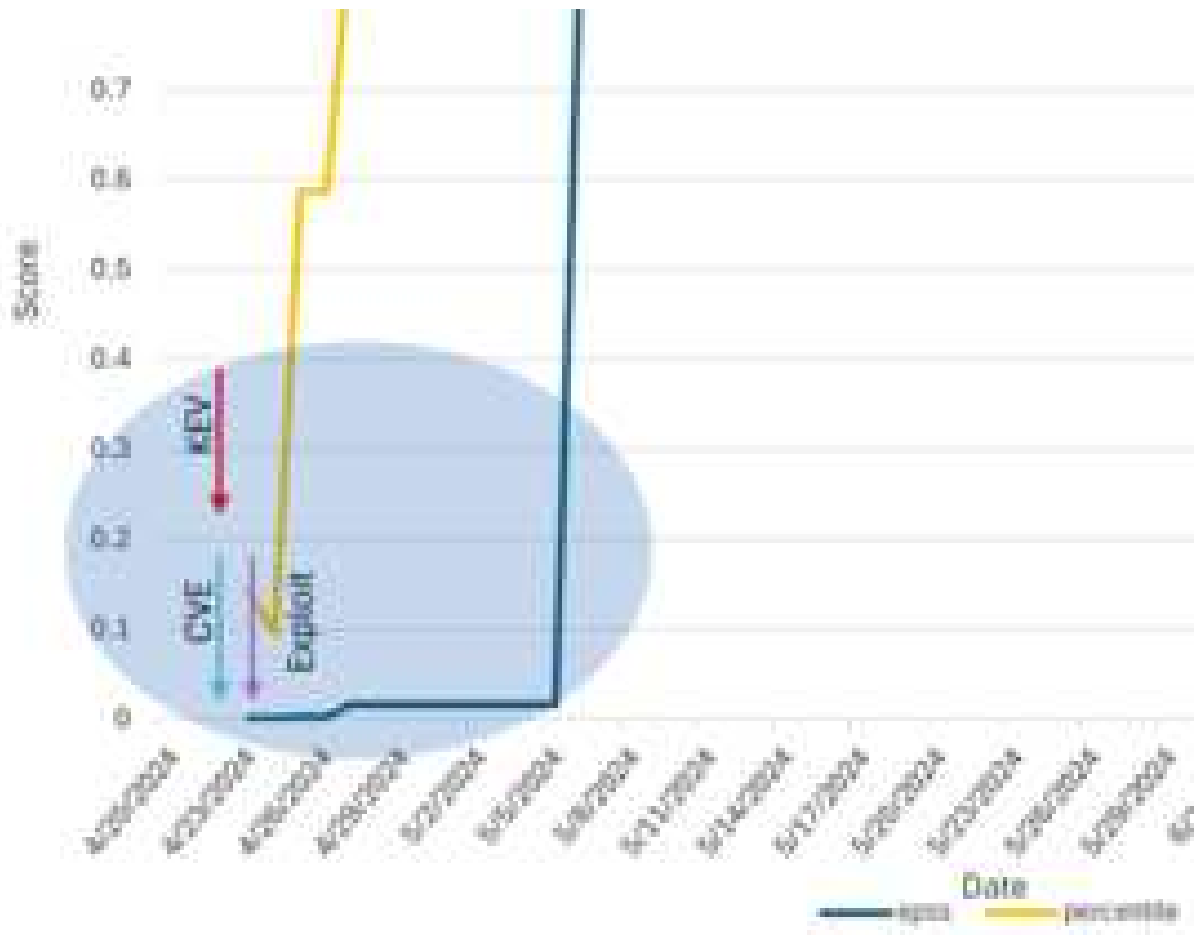

Fig. 9: CVE-2024-4040 Analysis

In most of the CVEs studied, exploit code was immediately available on the internet and several newsworthy articles were published shortly after the CVE was disclosed. Once a CVE is added to CISA KEV, it is known to have been already exploited. In such a case, the EPSS score is not reflecting that and often times significantly lags the CISA KEV catalog addition. However, since the data used in the modeling is not publicly available, it is difficult to assess what might be the reason for this significant lag and how to improve the model. A patch prioritization strategy that relied on EPSS would leave systems vulnerable during a period when exploit code was readily available and significant public disclosure of the vulnerability was available to allow a threat actor to quickly weaponize an attack against a target system. Additional research is needed in this area to better understand why scores are trailing significantly relative to the CISA KEV catalog entry.

### D. CVE-2023-29357 (MS Sharepoint) analysis

Analysis of Microsoft SharePoint vulnerability is incomplete due to lack of publicly available data to fully evaluate the EPSS scoring history. On one hand, the EPSS system had a high probability score prior to CISA KEV inclusion. However, that score dropped precipitously shortly to single digits before its CISA KEV catalog addition, and remained relatively low (below 36%) for about a month, eventually returning to the prior high probability score some months later. It is possible that the EPSS system accurately predicted the exploitation that resulted in the addition to CISA KEV. Prior to the EPSS high probability score in the EPSS system, this CVE spent nearly 5 months with an EPSS score near 0 despite a video demonstration of the exploitation at the Zero Day Initiative conference in March, 2023 [31]. This demonstration was later followed with a detailed blog of the exploit chain on 09/25/2023 [32]. It is difficult to know conclusively if that blog and related publicity were the cause in the sharp change in EPSS score, which occurred on 10/01/2023, or if there were observed exploit attempts that triggered this score change. At this time, there is insufficient publicly available data to make a determination. In order to complete the analysis for CVE-2023-29357, both CISA and the EPSS team would need to share additional details not currently available in the public domain.

### E. CVE-2023-42793 (JetBrains TeamCity) analysis

The JetBrains CVE was one of the better outcomes in terms of the EPSSv3 prediction scoring system as it correctly observed the exploitation of this vulnerability in the wild prior to the CVE being added to the CISA KEV catalog. On 09/06/2023, researchers from Sonar discovered a critical vulnerability in JetBrains TeamCity, which was later identified as CVE-2023-42793 [33]. This vulnerability in TeamCity could lead to authentication bypass and remote code execution on the server. This vulnerability has been observed being actively exploited in the wild and was added to CISA KEV on 10/04/2023 [34]. Since September 2023, Russian Foreign Intelligence Service (SVR)-affiliated cyber actors had been targeting servers hosting JetBrains TeamCity software that ultimately enabled them to bypass authorization and conduct arbitrary code execution on those compromised servers. A detailed retrospective, similar to the one done by FIRST.org on the Log4J vulnerability against the EPSSv2 system, should be done on this CVE with respect to the EPSSv3 scoring system to determine what worked well in this case.

### F. CVE-2023-27997 & CVE-2024-21762 (Fortinet FortiOS analysis)

These two CVEs appear to be significant outliers when compared to other remote access and perimeter security systems with similar vulnerabilities. Despite both of these CVEs being present in the CISA KEV catalog, neither have ever achieved a probability score above the 36% threshold that would result in prioritization of these vulnerabilities for remediation. In fact, CVE-2024-21762 has had an EPSS score near zero for the entire 6 months that the CVE has existed. There is evidence that this vulnerability may have been recently exploited based on vulnerability scan searches [35] [36] [37]. It is worth noting that the EPSS system uses Fortinet feeds as part of their modeling approach as discussed earlier and this might represent some sort of data blind spot for the system. Further detailed analysis of these two CVEs will be needed to understand why the EPSS scores have remained significantly lower than similar products from other vendors.

## Conclusion

EPSS Scoring system appears to be a trailing indicator of exploitation rather than a predictive system. The focus of the system is around probabilities of exploitation versus other vulnerabilities. In many cases the scoring accurately reflects the probability of exploitation prior to addition to the CISA KEV catalog, while in other cases it does not. Since the intent of EPSS is to help Threat Defenders prioritize which systems to patch, it is important that such a system has a very high efficacy in order to achieve that goal. The research findings and examples show that CVEs are often scored relatively lowly on the probability percentage at or around the time of entering into the CISA KEV catalog. In theory, the prediction system should have scored vulnerabilities higher prior to the addition to this catalog. Instead, these findings show that the scoring either coincides or trails the presence in the CISA KEV catalog.

The EPSS system authors clearly state it is less of a prediction system and more of a probability ecosystem of exploitation relative to other vulnerabilities [38]. That said, it would be reasonable to expect that if a vulnerability exists before being included in CISA KEV, then the probability system should have reflected it would likely have been exploited relative to other vulnerabilities at that same time. In general, the research does not support that conclusion.

The Exploit Prediction Scoring System (EPSS) name may be a bit misleading as it is less of a prediction system and more of a risk scoring system. In other words, what is the *risk* of exploitation of a given vulnerability versus other vulnerabilities. Most of the analyzed data showed that EPSS was less effective as a measurement than simply using the CISA KEV catalog directly. However, there were a few occasions where the EPSS system outperformed CISA KEV and should be considered as part of an overall risk mitigation strategy for vulnerabilities.

A defense in depth strategy is always a good practice and as the research showed, occasionally the EPSS system had a high probability score prior to a CVE being included in CISA KEV. This is a good example where the EPSS score would have helped prioritize a vulnerability that was being actively exploited in the wild before the CISA KEV publication of that vulnerability. By using CVSS, CISA KEV and EPSS in combination as part of a remediation strategy, a defender will be able to have the best possible outcome overall.

## Future Work

The EPSSv3 model is relatively new and needs more evaluation over the next couple of years. It would be useful for the EPSS creators to do a self evaluation of their latest model and provide more publicly available data so independent evaluation can be performed on their system. Additional research using the CISA KEV catalog as a gauge is also worthwhile as it provides a good control for this type of study. Of particular interest is to more closely analyze CVEs that had meaningful probability scoring changes before the CVE was added to the CISA KEV catalog to determine if it was a good prediction of the exploitation relative to other vulnerabilities. Another area of opportunity would be to explore additional machine learning algorithms and perhaps the application of Large Language Models (LLMs) using the data features available for the EPSSv3 system along with numerical importance scoring experiments. In order for these tests to be performed, it may require a larger set of features, more data or an LLM capable of operating with less information. Adding more threat feeds, including additional vendors to complement Fortinet, may prove to be beneficial, providing additional label data for the system. Additionally, threat sandbox vendors such as Joe Sandbox, Hybrid-Analysis or similar systems might provide different label data that would be useful for modeling.

## Appendix A:
## *ADDITIONAL CVE SAMPLE DATA*

### *A. CVE-2023-29357 (productivity software)*

Microsoft SharePoint Server contains a vulnerability that allows bypassing authentication, thus enabling the attacker to gain administrator privilege. Microsoft SharePoint on premise server is a popular platform for document collaboration at large organizations. Exploitation of this vulnerability can lead to compromise of the solution. This vulnerability was demonstrated at the Zero Day Initiative conference in March, 2023 [31] and a detailed analysis was provided on 09/25/2023 [32]. A Metasploit module was made available on 04/18/2024.

bgImg

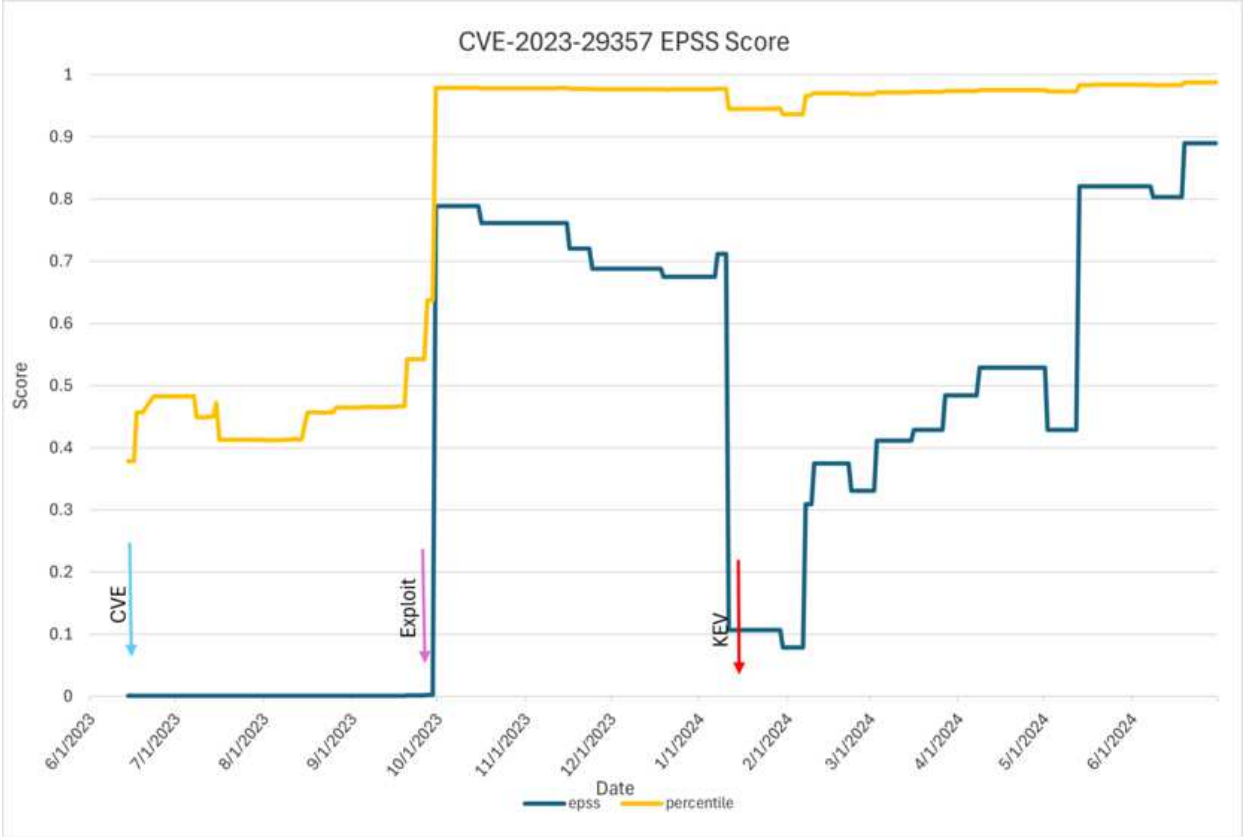

Fig. 10: CVE-2023-29357 EPSS Details

### *B. CVE-2023-34362 (secure FTP solution)*

MOVEit Transfer web application vulnerability that could allow an unauthenticated attacker to gain access to MOVEit Transfer's database. MOVEit Transfer is a managed file transfer (FTP) software solution widely used in enterprises. It is often exposed directly on the internet. A detailed analysis of the vulnerability was provided by Google Mandient on 06/02/2023. [39]. It was also known to have been used in a ransomware attack [40]. A Metasploit module was made available on 09/11/2023.

bgImg

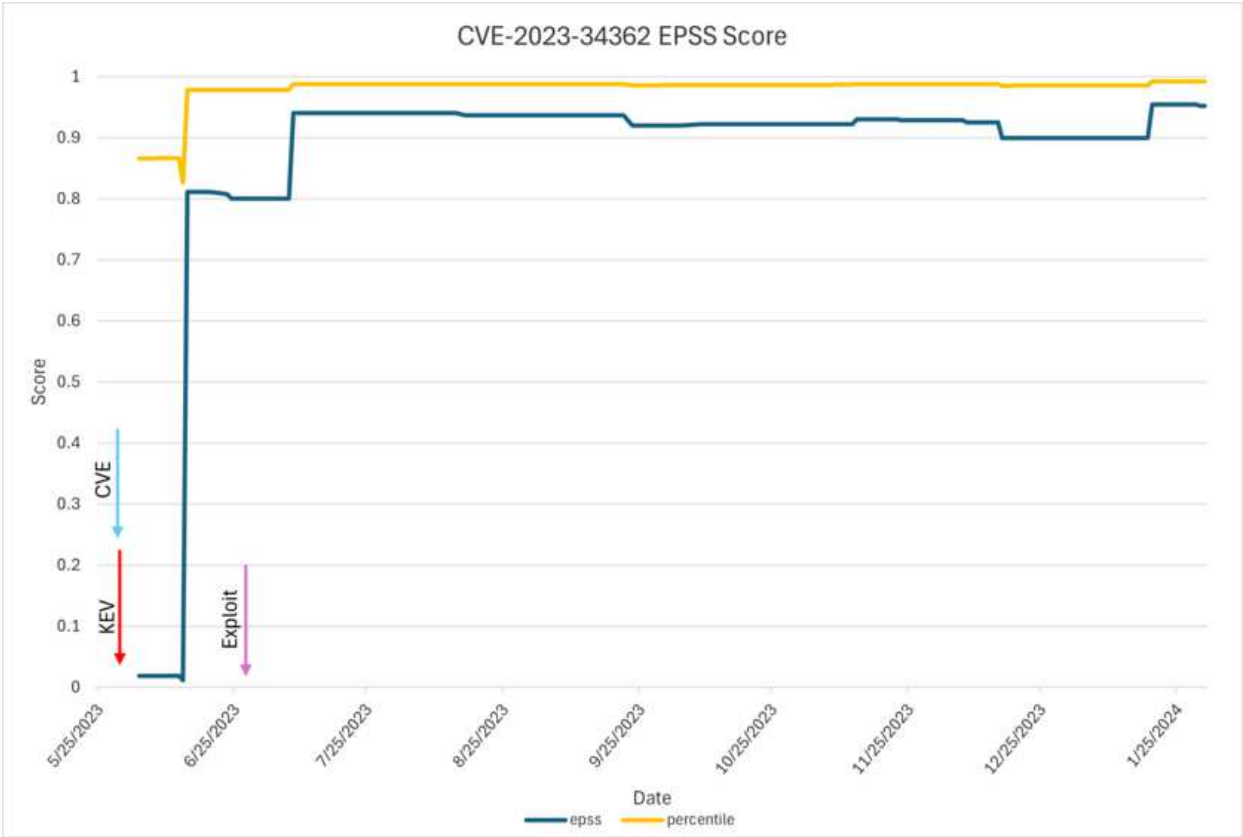

Fig. 11: CVE-2023-34362 EPSS Details

### *C. CVE-2023-35078 (enterprise mobility management)*

Ivanti Endpoint Manager Mobile (EMM) authentication bypass vulnerability. Ivanti EMM (*formerly MobileIron*) is a popular enterprise mobility management solution that is used to manage remote endpoints. A compromise of this system could allow a threat actor to deploy malicious software or configuration to an endpoint device. A detailed write up about this vulnerability from Palo Alto Unit42 was available on 07/28/2023, noting that as many as 5500 vulnerable systems were exposed to the internet [41]. This vulnerability is believed to have been used in a ransomware attack.

bgImg

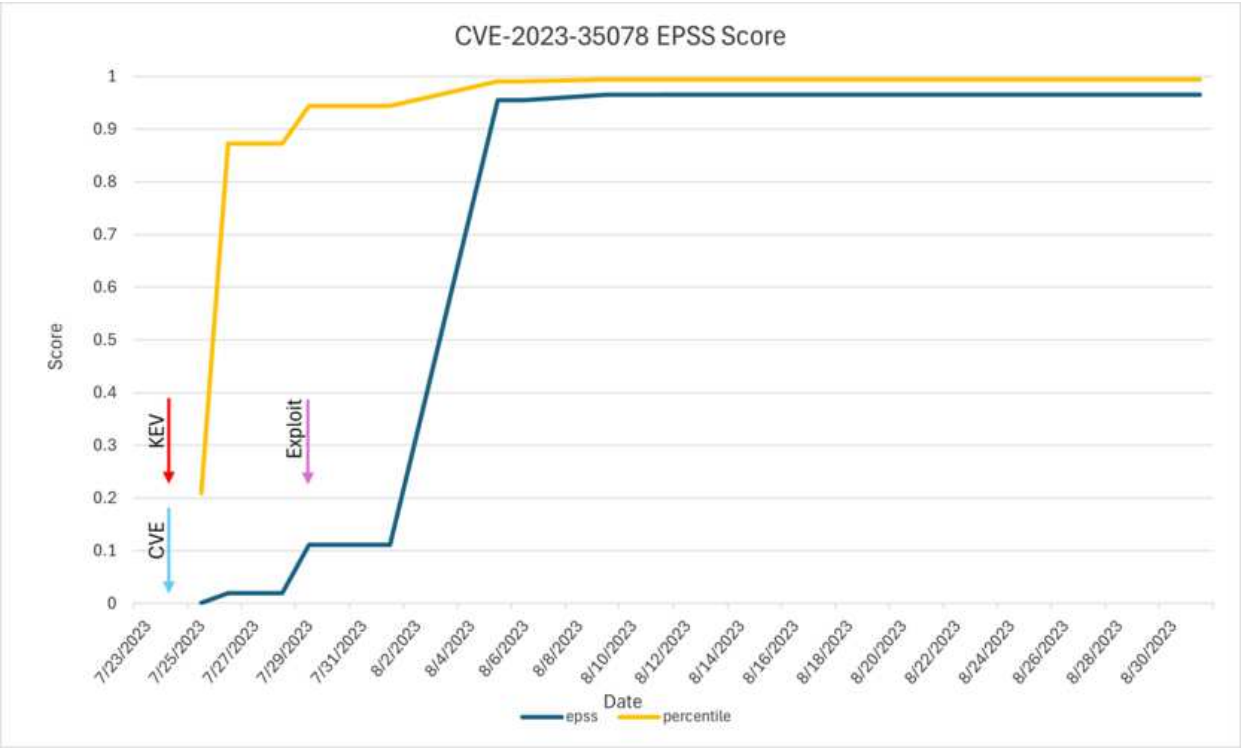

Fig. 12: CVE-2023-35078 EPSS Details

### *D. CVE-2023-42793 (supply chain software)*

JetBrains TeamCity authentication bypass vulnerability. TeamCity is a widely used open DevOps platform for continuous integration and deployment tooling (CI/CD). This vulnerability in TeamCity could lead to authentication bypass and remote code execution on the server. On 09/06/2023, researchers from Sonar discovered a critical vulnerability in JetBrains TeamCity, which was later identified as CVE-2023-42793 [33]. Since September 2023, Russian Foreign Intelligence Service (SVR)-affiliated cyber actors (also known as Advanced Persistent Threat 29 (APT 29), the Dukes, CozyBear, and NOBELIUM/Midnight Blizzard) have been targeting servers hosting JetBrains TeamCity software that ultimately enabled them to bypass authorization and conduct arbitrary code execution on the compromised server [34] [42]. The vulnerability was not added to the CISA KEV catalog until 10/04/2023, well after the first public exploit was available [43]. This vulnerability was known to have been used in a ransomware attack [44]. A Metasploit module was made available on 10/08/2023.

bgImg

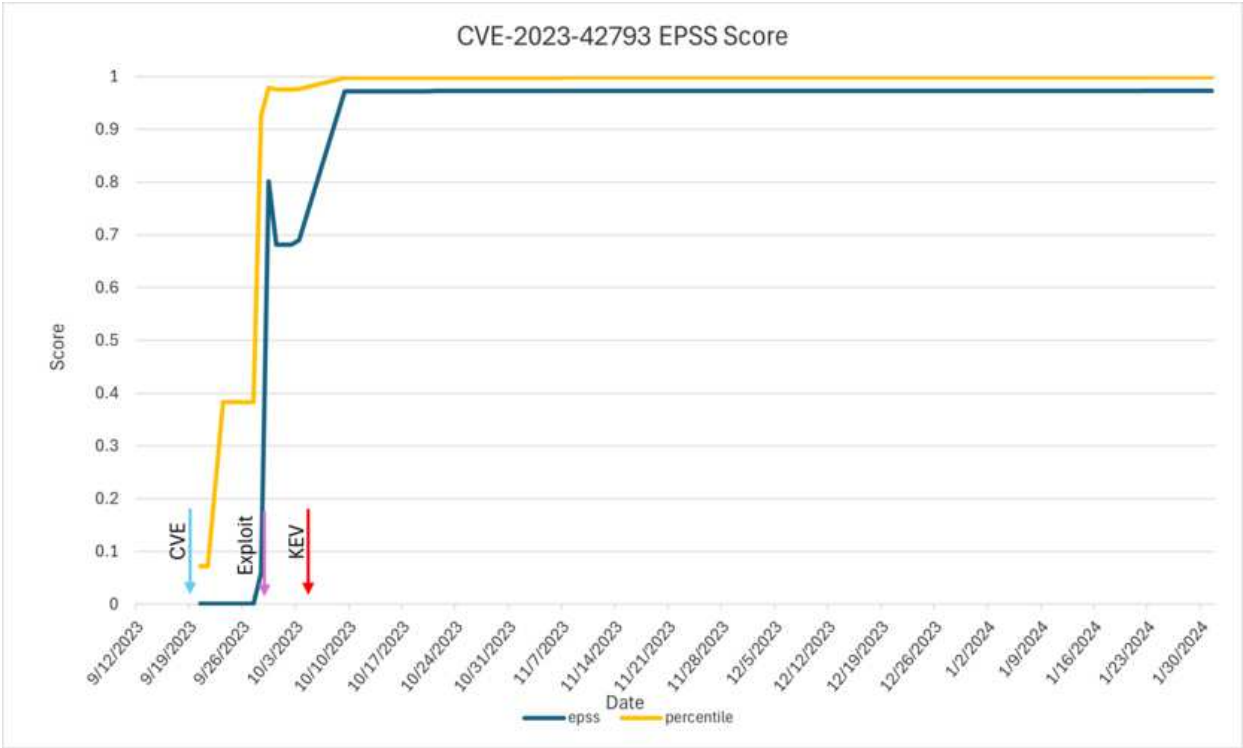

Fig. 13: CVE-2023-42793 EPSS Details

### *E. CVE-2023-46604 (Java messaging protocol library)*

Apache ActiveMQ deserialization vulnerability in OpenWire protocol. ActiveMQ is a multi-protocol messaging framework popular in Java programs. Because this is an API framework, many Java programs contain this vulnerability, very similar to how Log4j was also widely deployed in software. An attacker can compromise any system that is using the vulnerable version of this library.

A detailed write up about this vulnerability was published online on 12/11/2023 [45]. A public exploit was available the day after the CVE was released [46]. This was known to have been used in the *Godzilla* ransomware attack [47]. A Metasploit module was made available on 11/07/2023.

bgImg

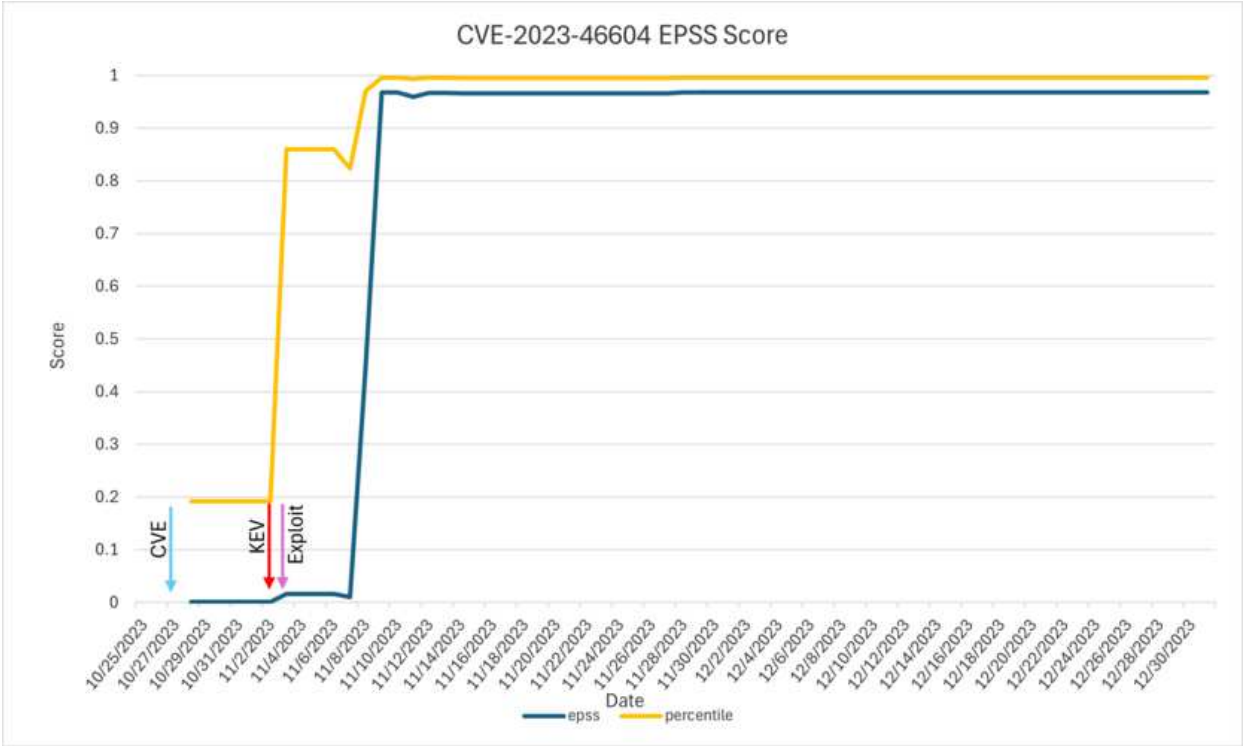

Fig. 14: CVE-2023-46604 EPSS Details

### *F. CVE-2023-48788 (enterprise mobility management)*

Fortinet FortiClient Endpoint Management Server (EMS) SQL injection vulnerability. FortiClient EMS is a popular enterprise mobility management solution that is used to managed remote endpoints. A compromise of this system could allow a threat actor to deploy malicious software or configuration to an endpoint device. A detailed write up, including a publicly available exploit was published on 03/18/2024 [48]. This vulnerability is known to have been exploited in a ransomware attack [49]. A Metasploit module was made available on 04/20/2023.

bgImg

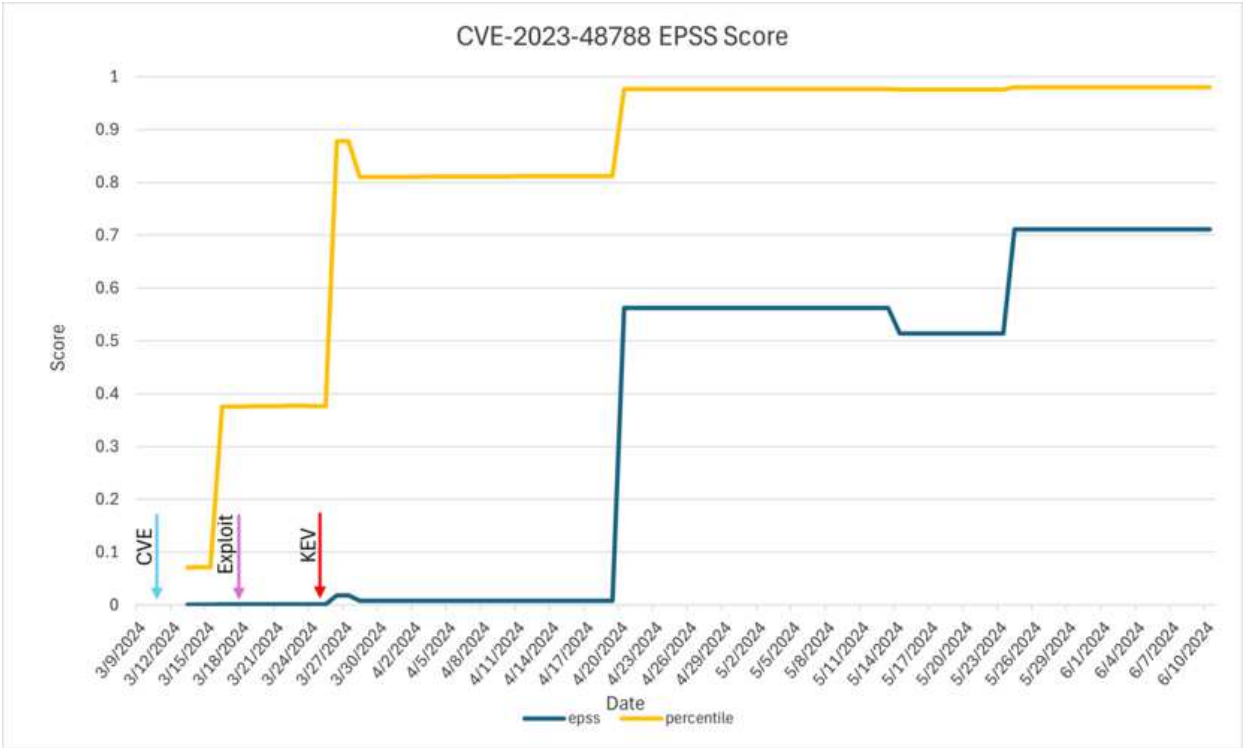

Fig. 15: CVE-2023-48788 EPSS Details

### *G. CVE-2024-21762 (remote access, perimeter security)*

Out-of-bounds write in Fortinet FortiOS that affects Fortinet appliances including remote access and network systems. Fortinet is a popular security appliance widely deployed in enterprises, often with direct exposure to the internet. A detailed write up on how to exploit the vulnerability titled *"Two Bytes is Plenty: FortiGate RCE with CVE-2024-21762"* was available on 03/15/2024 a few days after the CVE was originally published (03/09/24) [50].

bgImg

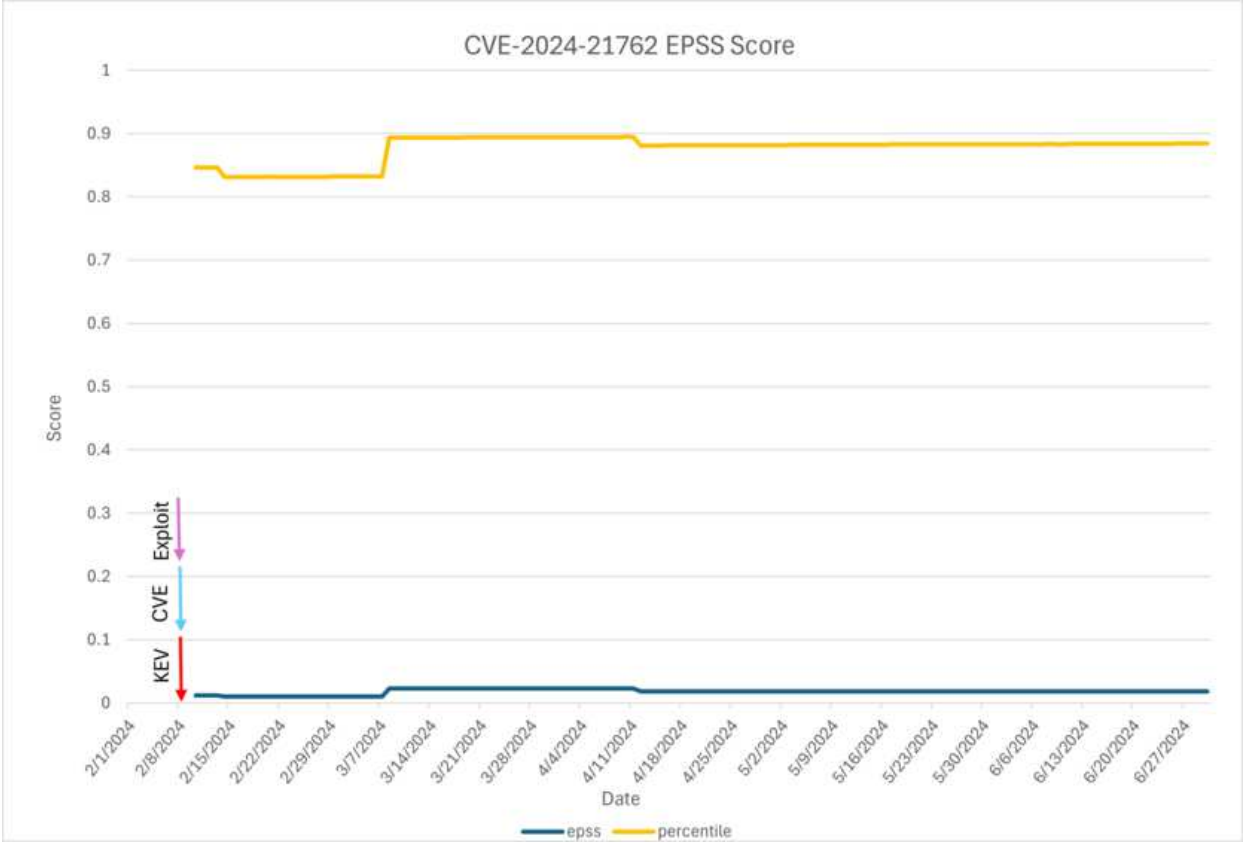

Fig. 16: CVE-2024-21762 EPSS Details

### *H. CVE-2024-4040 (remote file sharing)*

CrushFTP VFS Sandbox escape vulnerability. CrushFTP is a popular multi-protocol remote file sharing system, widely deployed in enterprises, typically with direct exposure to the internet. A sandbox escape can allow access to files on the underlying OS which can leak sensitive data or bypass authentication to execute code remotely leading to full system compromise. A detailed write up about this vulnerability was published online on 04/24/2024 [51]. A public exploit was available the day after the CVE was released [52]. A Metasploit module was made available on 05/20/2024.

bgImg

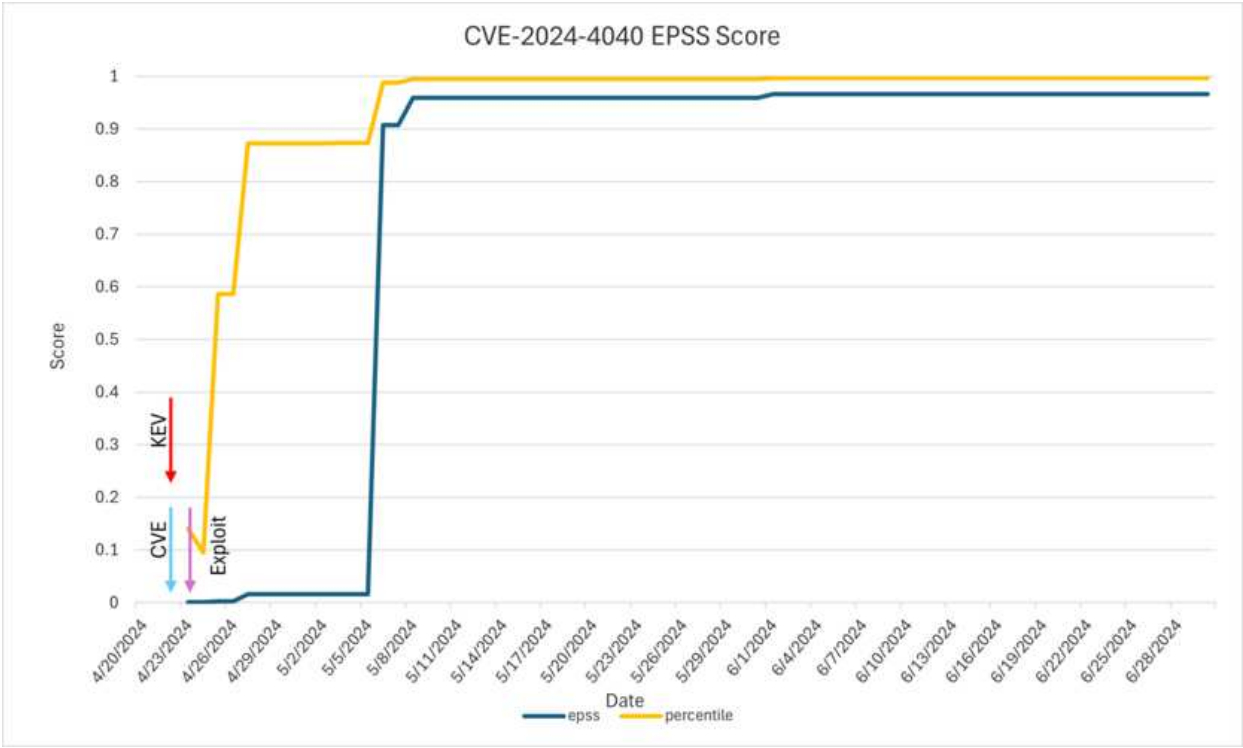

Fig. 17: CVE-2024-4040 EPSS Details

### I. *CVE-2024-4577 (web development software)*

PHP OS command injection vulnerability affecting PHP-CGI. PHP is a popular web development framework used on thousands of web servers on the internet. A vulnerability of this kind would allow a remote attacker to execute arbitrary code on a web server, potentially taking complete control of the system. A public exploit was available on 06/07/2024, prior to the creation of the CVE [53]. A detailed analysis of the vulnerability was also published a few days before the CVE on 06/06/2024 [54]. The vulnerability was used to distribute the *TellYouThePass* ransomware [55]. A Metasploit module was made available on 6/24/2024.

bgImg

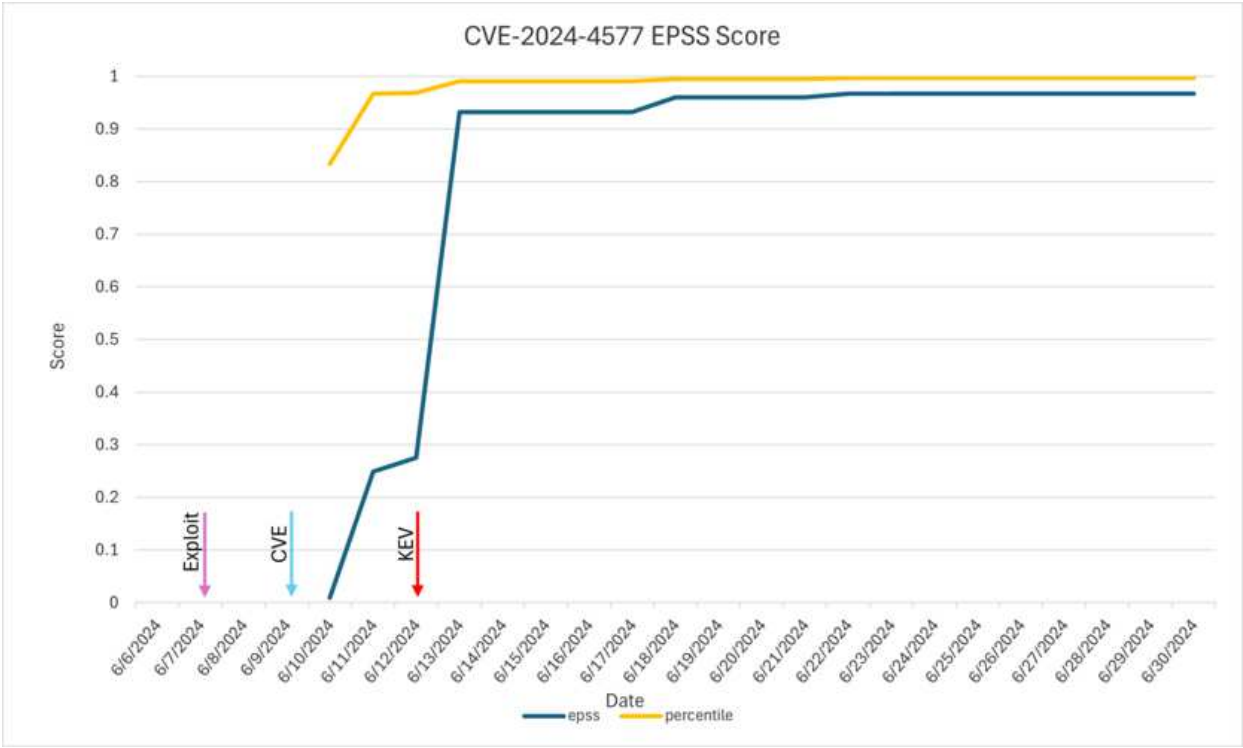

Fig. 18: CVE-2024-4577 EPSS Details